\documentclass[aps,prl,reprint,superscriptaddress,longbibliography,floatfix]{revtex4-2}
\usepackage[T1]{fontenc}
\usepackage{amsmath,amssymb,bm,graphicx,booktabs}
\usepackage[colorlinks=true,citecolor=blue,linkcolor=blue,urlcolor=blue]{hyperref}
\hypersetup{pdftitle={Magnon Chirality Selection at Equal Linewidths in Open Collinear Magnets},pdfauthor={Dinghui Wang, Lulu Liu}}
\newcommand{\kvec}{\mathbf k}
\newcommand{\Id}{\mathbb I}
\newcommand{\Tr}{\operatorname{Tr}}

\newcommand{\Dc}{\widetilde D}
\begin{document}
\title{Magnon Chirality Selection at Equal Linewidths in Open Collinear Magnets}
\author{Dinghui Wang}
\affiliation{School of Materials Science and Physics, China University of Mining and Technology, Xuzhou 221116, China}
\affiliation{Jiangsu Physical Science Research Center, Nanjing 210093, China}
\author{Lulu Liu}
\affiliation{School of Electronic Engineering, Nanjing Xiaozhuang University, Nanjing 211171, China}
\date{November 11, 2026}
\begin{abstract}
Loss can select one magnon chirality even when all resonances have the same linewidth. We construct a passive collinear mode network in which spatially correlated loss and coherent exchange give opposite angular momenta different survival probabilities. An initially unpolarized mixture reaches 26.9\% polarization with 29.9\% of the population remaining; an isospectral normal control gives none. A combined spin and crystal rotation reverses the signal and exchanges the exceptional contours reached at stronger loss. We organize the allowed pole pairings by the action of a non-Hermitian collinear spin-space little group on chirality and complex frequency. A second realization separates conjugate chirality sectors at opposite diagonal edges despite zero total winding. The two readouts distinguish constraints on poles from constraints on preparations, eigenvectors, and physical reservoirs.
\end{abstract}
\maketitle
Open quantum and classical wave systems exhibit dynamical phenomena that have no Hermitian counterpart, including eigenvector coalescence at exceptional points, complex spectral topology, and boundary-sensitive mode accumulation~\cite{Hatano1996,Heiss2012,Miri2019,Ozdemir2019,Bergholtz2021,Ding2022,Okuma2023,Ashida2020,ZhangReview2022,Gong2018}. Their effective non-Hermitian generators contain information not only in complex poles but also in the relations between right and left eigenvectors and in the couplings to physical reservoirs. Symmetry constrains this information through operations that can involve transposition or complex conjugation rather than ordinary unitary similarity~\cite{Kawabata2019,Zhang2023,Sa2023}. A central task is therefore to connect these spectral and mode-space constraints with measurable dynamics.

Magnons provide a natural setting for this connection because their angular momentum, band degeneracies, and topology are tightly controlled by magnetic symmetry. In a closed magnet the combination of crystal and independent spin rotations determines these properties through spin-space groups~\cite{Brinkman1966,Corticelli2022,Xiao2024,Jiang2024,Chen2025,Fishman2022,Sandratskii2025}. Dissipation adds another layer of structure: magnetic exceptional points, nonreciprocal spin dynamics, and boundary accumulation have been predicted or observed in a range of magnonic platforms~\cite{Zhang2017,Liu2019,McClarty2019,Tserkovnyak2020,Galda2019,ZhangYou2019,Wang2021,Deng2022,Grigoryan2022,Sadovnikov2022,Wang2023,Wittrock2024,Gunnink2024}. Extending collinear spin-space symmetry to this open setting requires a description of both the dynamical generator and the reservoir channels.

The chirality of a magnon offers a particularly direct probe of that description. In itinerant altermagnets, coupling to electronic excitations broadens the two angular-momentum flavors differently~\cite{Costa2025,Issing2026}. In a metal-capped ferromagnet, chiral damping instead distinguishes incident propagation directions and changes transmission across a barrier~\cite{Ye2024}. These effects are identified through pole frequencies, linewidths, or directional response. The present comparison is narrower: opposite angular momenta at fixed momentum have exactly the same pole linewidths at the operating point.

Two magnon pulses prepared with opposite angular momenta can nevertheless leave different populations after a common delay. Measuring all resonance frequencies and linewidths need not resolve this difference: a system with precisely the same poles but orthogonal modes can show no population contrast at all. This preparation-dependent response belongs to the broader physics of nonnormal dynamics, where transient behavior depends on the propagator and the initial state as well as on the eigenvalues~\cite{Makris2014,OkumaNakagawa2022,Makris2021}. The resulting symmetry problem is to distinguish which properties of an open collinear magnet follow from the pairing of complex frequencies and which require the action of symmetry on the modes, loss channels, and preparation.

We give a passive realization in which the measured polarization distinguishes two systems that even complete pole spectroscopy cannot tell apart. Our starting point is the spin-space representation of transverse modes: axial rotations assign opposite charges, and a combined spin and crystal operation can exchange them~\cite{Chen2025}. We determine how the loss channels must transform to preserve this operation and identify the constraints imposed by an additional symmetry of their centered drift that is not a symmetry of the bath. Five images of the generalized little group summarize the resulting pole constraints. The population measurement probes information discarded by this reduction. We also show how to read that information spatially by comparing opposite edges at a single parallel momentum.

\begin{figure*}[!t]\centering\includegraphics[width=\textwidth]{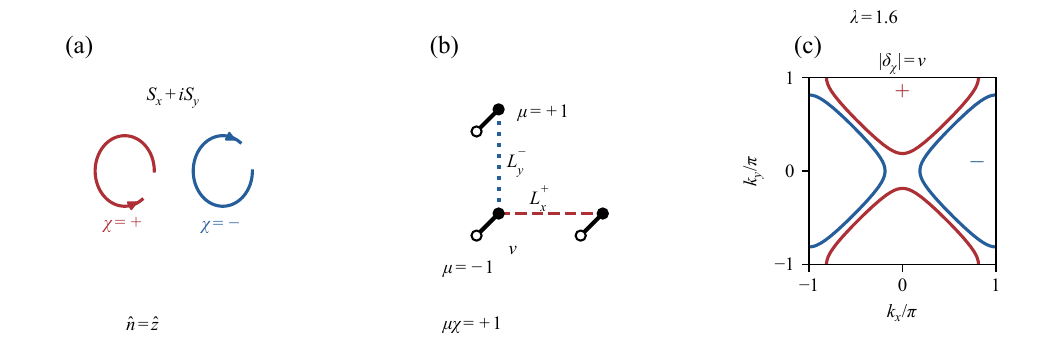}\caption{Mode schematic and calculated contours. (a) Opposite senses of transverse precession label two positive-frequency chiralities of a collinear magnet. (b) Each pair of modes (filled/open) has coherent coupling $v$. Dashed and dotted bonds denote collective losses with opposite relative signs at fixed $\mu\chi=+1$. (c) At $\lambda=1.6$ a quarter turn exchanges the $\chi=+$ (red) and $-$ (blue) exceptional contours. Each coalescence occurs within one chirality.}\label{net}\end{figure*}

\emph{What the group acts on.}---Fix a collinear configuration $\mathbf m_j=m_j\hat{\mathbf n}$ and choose canonical positive-frequency modes. Set $\hbar=1$. Within Markovian open-system dynamics~\cite{Lindblad1976,Gorini1976,Prosen2008,Albert2014,Minganti2019}, a number-conserving quadratic Hamiltonian $b^\dagger h b$ and linear vacuum-loss channels $L_\nu=\ell_\nu b$ give the first-moment equation
\begin{equation}
 i\partial_t\langle b\rangle=D\langle b\rangle,\quad D=h-i\Gamma,\quad \Gamma=\ell^\dagger\ell/2\succeq0 .
\end{equation}
Thus every amplitude contracts in norm, including when the eigenvectors of $D$ are nonorthogonal. Assume the environment as well as the magnet respects axial rotations $b_\chi\mapsto e^{i\chi\theta}b_\chi$. Then $D=D_+\oplus D_-$, with $\chi=\pm1$ denoting opposite angular momenta. A loss jump can transfer angular momentum to the bath and still respect this covariance: its axial phase cancels from its dissipator. This assumption is stronger than collinearity of the static order.

A spin-space operation $g=\{R_s\Vert R|\mathbf t\}$ preserving the magnetic background acts on momentum by $q_g$ and on the modes by a sewing matrix $U_g$. Site permutations, axial charges and translation phases are fixed by the chosen representation. Extend it by a matrix tag $\alpha_g\in\{1,T,*,\dagger\}$, allowing specified internal involutions. With one symmetry-invariant decay center $\gamma_c$ for both chiralities, define $\Dc=D+i\gamma_c\Id$. The resulting non-Hermitian collinear spin-space group acts by
\begin{equation}
 U_g(\kvec)\alpha_g[\Dc(q_g^{-1}\kvec)]U_g^{-1}(\kvec)=\Dc(\kvec).
\end{equation}
The common shift specifies where to reflect a complex frequency. It is never removed from the physical propagator. Multiplying two operations composes the tags, but the sewing product is twisted: transposing $VXV^{-1}$ puts $(V^T)^{-1}$, rather than $V^T$, on the left of $X^T$. The resulting multiplication law is given in End Matter. In particular, equivalent descriptions must transform the sewing matrices together with the dynamical matrix. Treating all invertible similarities as crystal operations would discard the physical embedding and give a different group for arbitrary choices of coordinates.

Three distinctions are essential. First, a transpose exchanges right and left eigenvector equations without conjugating the eigenvalue. Second, a bosonic Nambu drift can be non-Hermitian even in a closed magnet: canonical changes of basis preserve the commutator metric $J$, and the negative-frequency partner describes the adjoint mode, not the other positive-frequency chirality. Third, a centered drift relation need not be a symmetry of the reservoir. A physical unitary operation must preserve $h$ and can mix the rate-weighted loss channels by a unitary transformation. This stronger condition also fixes the transformed noise. We impose it for the combined spin and crystal rotations below, while the additional conjugation relations need hold only for the centered drift.

\emph{Five spectral constraints.}---At fixed momentum keep the operations that return that momentum modulo a reciprocal lattice vector and preserve the frequency sector. On its multiset of centered poles, each operation has just two signs: $s$ records whether it preserves or exchanges the chirality blocks, and $c$ whether it leaves a pole unchanged or takes its complex conjugate. Identity and transpose have $c=+1$; conjugation and adjoint have $c=-1$. Both signs multiply under composition. Write $a=(-,+)$ and $b=(+,-)$. The image is a subgroup of the Klein four-group. Besides the trivial and full groups there are exactly three subgroups, each generated by one of $a,b,ab$. Any two distinct nonidentity elements force the third.

The trivial image ($\mathsf S_0$) enforces no pairing. The image generated by $a$ ($\mathsf D_e$) gives equal spectra in opposite chiralities. The image generated by $b$ ($\mathsf S_c$) closes each spectrum separately under conjugation, leaving the sectors free to differ. The image generated by $ab$ ($\mathsf C$) pairs one chirality with the conjugate of the other. The full image ($\mathsf D_c$) requires both equality and conjugate closure. Equality includes algebraic multiplicities; it need not identify a globally continuous band label across an exceptional point. These five little-group images classify the enforced pairings of centered poles in the two chirality sectors. An accidental equality of measured poles does not establish the corresponding symmetry.

Table~\ref{classes} gives a useful spectroscopic distinction. In $\mathsf S_c$ two individually real centered poles can acquire opposite imaginary parts without breaking the symmetry: the multiset remains closed. In $\mathsf D_e$, by contrast, each pole must have an opposite-chirality partner of the same frequency and linewidth. Perturbations preserving the operation distinguish enforced pairing from accidental coincidence.
\begin{table}[tb]
\caption{Little-group images and their mandatory constraints on the centered spectrum $S_\chi$. The star conjugates the entire multiset.}\label{classes}
\begin{ruledtabular}\begin{tabular}{ccc}
Class & Generators of image & Constraint\\
$\mathsf S_0$ & none & none\\
$\mathsf D_e$ & $a$ & $S_+=S_-$\\
$\mathsf S_c$ & $b$ & $S_\chi=S_\chi^*$\\
$\mathsf C$ & $ab$ & $S_+=S_-^*$\\
$\mathsf D_c$ & $a,b$ & $S_+=S_-=S_+^*$
\end{tabular}\end{ruledtabular}
\end{table}

Across different momenta the same signs distinguish frequency from lifetime textures. For consistently matched branches away from exchanges, the real part of the chirality difference transforms by $s$, the imaginary part by $sc$. A rotation that exchanges chirality and conjugates the frequency can therefore reverse a real splitting while preserving a linewidth difference. Inside a multiplet the ordinary unitary little subgroup adds a separate restriction. If an irreducible representation of dimension $d$ occurs $m$ times, its commuting drift is $\Id_d\otimes B_m$ by Schur's lemma. Counting constraints on eigenvalue coalescence in a general multiband matrix~\cite{Sayyad2022} and locating its Jordan chains in a given spin representation answer different questions. A single copy enforces an ordinary degeneracy, even for a complex pole. A Jordan chain of length $r$ in $B_m$ gives $d$ chains of length $r$, never a single chain of length $dr$. Extra spectral operations can further restrict $B_m$; multiple copies provide room for a defect without guaranteeing one.

\begin{figure}[!t]\centering\includegraphics[width=\columnwidth]{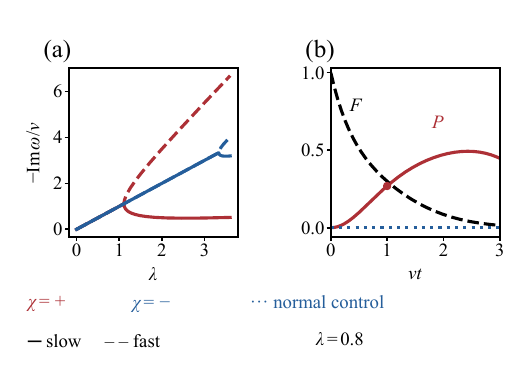}\caption{Chirality selection at $f_d=1$. (a) Slow (solid) and fast (dashed) pole decays split at different losses for $+$ (red) and $-$ (blue). (b) At $\lambda=0.8$ all poles have the same linewidth. Each chirality starts with $C_\chi=\mathbb I/2$. Differential contraction produces polarization $P$ as its surviving fraction $F$ falls. The dotted line is the isospectral normal control; the dot marks $vt=1$.}\label{pop}\end{figure}

\emph{A reservoir that selects chirality.}---Consider two internal modes per chirality with coherent coupling $v>0$. Their chirality-independent Hamiltonian is $\omega_0\Id+v\sigma_x$. Local losses and interference between loss amplitudes on neighboring cells produce
\begin{equation}
 \Gamma_\chi=\lambda\gamma_b\Id+\delta_\chi\sigma_z,\qquad
 \delta_\chi=\lambda(d_0+\chi d_1 f_d),
\end{equation}
where $f_d=(\cos k_x-\cos k_y)/2$. Increasing the bath strength $\lambda\ge0$ leaves the conservative magnet fixed. Figure~\ref{net} shows the relative phases that generate the two cosine terms: what interferes constructively along one axis interferes destructively along the other. With $\gamma_b\ge |d_0|+d_1$ and $d_1\ge0$, every channel has a nonnegative rate (End Matter). A quarter turn sends $f_d$ to $-f_d$; exchanging chirality at the same time maps the full set of jumps into itself.

In contrast, $\sigma_x\Dc_\chi^*\sigma_x=\Dc_\chi$ at $\gamma_c=\lambda\gamma_b$ implements the $\mathsf S_c$ spectral relation but exchanges unequal internal losses. At fixed chirality the square-root problem is the dissipative spin dimer of Ref.~\cite{Tserkovnyak2020} in its reactive-coupling, differential-loss limit. Here the differential loss is tied to a bond orientation and an axial charge. Its effect is visible in the poles $\omega_{\chi,\pm}=\omega_0-i\lambda\gamma_b\pm\sqrt{v^2-\delta_\chi^2}$. The discriminant is real. Thus coalescence takes one real condition, $|\delta_\chi|=v$, giving a contour in two-dimensional momentum space. There the nonzero traceless block squares to zero: the two internal eigenvectors of \emph{one} chirality coalesce. Beyond the contour the conjugate centered poles have different lifetimes. A real mass $m\sigma_z$ preserves the spin-space quarter turn but breaks centered conjugation. It adds $m^2-2im\delta_\chi$ to the discriminant and removes the contour for $m\ne0$. The rotation relates two contours; conjugation accounts for their codimension.

Take $(d_0,d_1,\gamma_b,\omega_0)/v=(0.6,0.3,1,3)$. At $f_d=1$ the two thresholds are $10/9$ and $10/3$ [Fig.~\ref{pop}(a)]. On the rotation-related momentum the order is reversed. At $f_d=0$ both sectors reach threshold together at $5/3$. On the exchange-invariant diagonal the unitary little group enforces a chirality doublet and there are two internal copies. The four coincident poles are two second-order exceptional blocks, as the multiplicity rule predicts.

Below the first threshold all four linewidths coincide, although the square root depends on chirality. To probe what spectroscopy misses, prepare each sector in the same incoherent internal mixture $C_\chi(0)=\Id/2$ at selected momentum, and switch off the drive. Its surviving occupation is $S_\chi=\Tr(U_\chi U_\chi^\dagger)/2$ with $U_\chi=\exp(-iD_\chi t)$. For $\Omega_\chi^2=v^2-\delta_\chi^2>0$ this gives
\begin{equation}
 S_\chi=e^{-2\lambda\gamma_b t}\left[1+2\delta_\chi^2
 \frac{\sin^2(\Omega_\chi t)}{\Omega_\chi^2}\right].
\end{equation}
The common exponential cancels in $P=(S_+-S_-)/(S_++S_-)$. The bracket depends on how the two internal decay amplitudes interfere during coherent exchange. Equivalently, evolution is nonnormal because the coherent coupling and the differential loss do not commute. A normal control with the same poles replaces the traceless block by $\Omega_\chi\sigma_z$. Its propagation is unitary up to the common attenuation, and $P$ vanishes identically. Thus different real frequencies alone cannot explain the effect. Nonnormal power dynamics have been diagnosed using the largest singular value of a propagator, optimized over input states~\cite{Makris2014,OkumaNakagawa2022,Makris2021}. Our preparation instead weights both squared singular values equally in each sector. None can exceed unity in this loss-only realization.

There are two different orders of onset. Expanding the upper pole in weak loss gives a real chirality splitting $-2\lambda^2d_0d_1f_d/v+O(\lambda^4)$, although the coherent Hamiltonian contains no such term. Expanding the polarization at short time gives $P=4\lambda^2d_0d_1 f_dt^2+O(t^4)$. More generally, for an internal mixture $\Id/m$ the initial slope of $S$ measures the mean eigenvalue of $-2\Gamma$, and its curvature measures $4\Tr(\Gamma^2)/m$ (End Matter). Equal mean loss cancels the slope in the chirality difference; unequal loss variances can survive in the curvature. Equal mean loss alone need not give equal pole linewidths. The real discriminant below threshold supplies that additional condition here. A uniform offset $d_0$ and a sign-changing component $d_1 f_d$ are both needed: removing either makes $\delta_+^2=\delta_-^2$, eliminating the contrast exactly. Resolving a linewidth splitting thus requires crossing a finite threshold, whereas a time-resolved preparation can detect chirality selection below it.

At $\lambda=0.8$, $f_d=1$ and $vt=1$, the polarization reaches 0.269 while the surviving fraction $F=(S_++S_-)/2$ is 0.299. Both populations decrease: this is a selection among surviving excitations, without gain. A quarter turn reverses the polarization, so an equally weighted sum over rotation partners cancels. For a common Markov bath occupation the same signal applies to the excess above the stationary background; a thermal background included in the measured intensity dilutes it. These preparation and readout conditions distinguish the transient from a spontaneous polarization of the vacuum.

\begin{figure}[!t]\centering\includegraphics[width=\columnwidth]{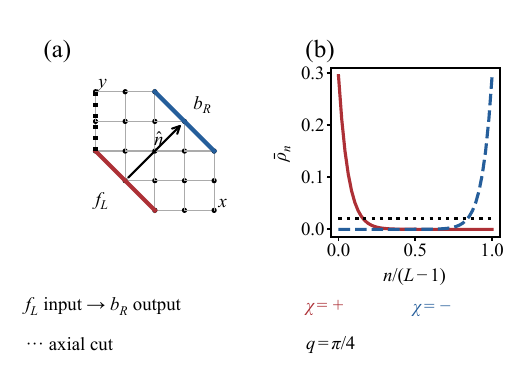}\caption{Geometry selects the response. (a) Plan view of diagonal layer edges (colored) and an axial edge (dotted); the arrow connects input and output layers. (b) At $q=\pi/4$, individually normalized right modes on 48 layers accumulate at opposite ends ($+$ red solid, $-$ blue dashed); the axial control is extended (dotted). These densities are distinct from the driven port intensities.}\label{edge}\end{figure}

\emph{A boundary that resolves cancelling windings.}---One mode per chirality suffices for the second example. Keep the coherent nearest-neighbor coupling $t$ and take
\begin{align}
 D_\chi={}&\omega_s+2t(\cos k_x+\cos k_y)\nonumber\\
 &-i[\gamma+2\chi a(\cos k_x-\cos k_y)] .
\end{align}
Here $\gamma>4a>0$, $\omega_s>4|t|$; local and bond losses again realize the whole drift. Centering at $\gamma$ pairs opposite chiralities by conjugation ($\mathsf C$). For a loop with a point gap about $z_0=\omega_s-i\gamma$, conjugation reverses the phase winding of $D_\chi-z_0$. The total winding consequently vanishes on the diagonal loop considered below. An axial symmetry still separates the two contributions; their cancellation in the determinant does not require either contribution to vanish. Geometry-dependent skin effects occur quite generally in higher dimensions~\cite{Zhang2022}. The wider connections among point-gap topology, non-Bloch bands, open-system damping, and boundary accumulation are well established~\cite{Yao2018,Kunst2018,Yokomizo2019,Okuma2020,ZhangWinding2020,Lee2019,Song2019,Kawabata2020,Okugawa2021,Claes2021}; here the separation by chirality fixes the paired response of the two edges.

For a strip normal to $x$, varying $k_x$ at fixed $k_y$ traces a line segment twice. The two hopping directions are equal and the open-strip modes are extended standing waves. Rotate the cut so that $k_x=p+q$, $k_y=p-q$, with $p$ normal and $q$ conserved parallel to the edge. The periodic spectrum relative to $z_0$ now traces the ellipse $4t\cos q\cos p+4i\chi a\sin q\sin p$. At $q=\pi/4$ the two ellipses coincide as sets but are traversed in opposite directions. At $q=0$ the point gap closes, so this argument cannot be extended continuously to every parallel momentum.

One can predict the allowed winding comparison before diagonalizing the strip. Define $w_\chi$ by the change of argument of $\det(D_\chi-z_0)$ divided by $2\pi$ around the normal-momentum loop. A symmetry reversing the loop changes this integer's sign; conjugating the centered spectrum also changes its sign, provided the reference point maps to the partner reference point. For a loop mapped to itself and a reference invariant in centered coordinates these two signs multiply. In particular, opposite-sector conjugation on an orientation-preserving loop allows equal and opposite windings. It does not force either integer to be nonzero. The explicit ellipse establishes nonzero winding; the axial line segment provides the control in which both sectors are trivial for any reference outside the segment. Its center lies on the segment and cannot itself be used as a gapped reference there.

In reciprocal skin systems the momentum slices at $q$ and $-q$ can already localize on opposite boundaries~\cite{Hofmann2020}. Here both chiralities are available at the \emph{same} $q$, allowing one to compare their responses with the same source geometry and frequency. In the honeycomb magnetic model of Ref.~\cite{Deng2022} spin-orbit coupling and damping open spectral area; in the present model the relative phases of loss channels supply the needed independent imaginary dispersion. The open strip makes the orientation of this loop observable. Its upper and lower hopping elements are $T_R=2t\cos q+2\chi a\sin q$ and $T_L=2t\cos q-2\chi a\sin q$. For positive hoppings the right-mode amplitude acquires an envelope $(T_L/T_R)^{n/2}$ multiplying a standing wave. Changing chirality inverts the envelope but leaves the product of hoppings, and hence all open-strip eigenvalues, unchanged. At $a/t=0.2$ and $q=\pi/4$ the amplitude decay length is 4.93 layers. Figure~\ref{edge} compares the mean right-eigenmode density for the two cuts, with each mode normalized separately. To predict the signal from a localized drive, we use the strip Green function.

At each edge we retain the onsite loss contribution of each cut bond. We drive the left endpoint equally in the two uncoupled chirality sectors and read the right endpoint. Transposition exchanges the two strip matrices, giving identical Green-function denominators. Their end-to-end numerators instead contain opposite hopping products. The output intensity ratio is exactly $|T_{+,L}/T_{+,R}|^{2(L-1)}$. For $L=12$, $\omega_d=\omega_s=6t$ and $\gamma=t$, it equals $(2/3)^{22}$: the output polarization is $-0.99973$. The corresponding squared end-to-end susceptibilities are $2.78\times10^{-6}$ for positive chirality and $2.08\times10^{-2}$ for negative chirality, in units of $t^{-2}$. Reversing the ports or selecting $-q$ exchanges the preferred chirality.

These examples suggest two complementary measurements on the same conceptual footing. One can reverse the time-resolved signal by rotating the selected momentum, and the spatial signal by exchanging the ports at fixed parallel momentum. In both tests the projected loss channels determine how the prepared modes reach the detector. Correlated loss, familiar from reservoir engineering~\cite{Metelmann2015,Fang2017,WangClerk2023,Cao2020}, supplies a concrete design variable: its relative bond phases determine the momentum parity of the loss contrast without changing the conservative hopping. Related nonreciprocal lattice dynamics also underlies directional amplification and sensing~\cite{McDonald2020}. The group fixes which poles must be partners, its representations determine where eigenvectors can merge, and the reservoir and the preparation determine what remains at a detector. This separation makes the rules applicable to a specified collinear magnet without assuming a universal damping mechanism.
\begin{acknowledgments}
We thank Yuting Yang for helpful discussions.
Dinghui Wang acknowledges support from the National Natural Science Foundation of China (Grant No.~12604376), the Fundamental Research Funds for the Central Universities (Grant No.~2024QN11046), and the Natural Science Foundation of Jiangsu Province (Grant No.~BK20233001).
Lulu Liu acknowledges support from the Basic Research Program of Jiangsu (Grant No.~SBK20260202315).
\end{acknowledgments}

\clearpage
\section*{End Matter}
\emph{Composition and admissible symmetry.}---Let $F_{U,\alpha}X=U\alpha(X)U^{-1}$. The composition $F_{U,\alpha}F_{V,\beta}$ has tag $\alpha\beta$ and sewing matrix $U\theta_\alpha(V)$, with
\begin{align}
 \theta_1(V)&=V,&\theta_*(V)&=V^*,\nonumber\\
 \theta_T(V)&=(V^T)^{-1},&\theta_\dagger(V)&=(V^\dagger)^{-1}.
\end{align}
Indeed, transposition and adjunction reverse the three factors of $VXV^{-1}$. With momentum restored the product is $U_g(\kvec)\theta_{\alpha_g}[U_h(q_g^{-1}\kvec)]$, up to the scalar factor system of the Bloch representation. The action is associative since $\theta_\alpha\theta_\beta=\theta_{\alpha\beta}$. The kernel of the tag map consists of ordinary similarities; its image records matrix operations. Its image is distinct from the chirality image in the Letter: transpose and identity can act identically on poles while treating eigenvectors differently.

For a canonical Nambu basis the metric is $J=\operatorname{diag}(\Id,-\Id)$ and a physical basis transformation obeys $TJT^\dagger=J$. The number-conserving completion of $D(\kvec)$ has diagonal blocks $D(\kvec)$ and $-D^*(-\kvec)$, whose paired poles share the same decay sign. This minus sign is the operator redundancy of the completion. Including it in the sign-preserving little-group test would count negative-frequency partners as extra positive-frequency chiralities.

\emph{Constructing the two baths.}---For the first model fix a cell $r$, chirality $\chi$ and internal label $\mu=\pm1$. In the following expressions $b$ denotes the corresponding annihilator and $b_x,b_y$ its translations by one lattice constant. Choose
\begin{align}
 L_0&=\sqrt{2\lambda(\gamma_b+\mu d_0-d_1)}\,b,\nonumber\\
 L_x&=\sqrt{\lambda d_1/2}(b+\mu\chi b_x),\nonumber\\
 L_y&=\sqrt{\lambda d_1/2}(b-\mu\chi b_y).
\end{align}
Their contributions to $\ell^\dagger\ell/2$ are the local loss and $\lambda d_1(1+\mu\chi\cos k_x)/2$, $\lambda d_1(1-\mu\chi\cos k_y)/2$. Summation reproduces the loss in the Letter. Under a quarter turn and chirality exchange the two bond channels map into one another, up to a translation and a phase. This is a symmetry of the channel sum and not just a relation between poles.

For the second model use $L_0=\sqrt{2(\gamma-4a)}b$, $L_x=\sqrt{2a}(b+\chi b_x)$ and $L_y=\sqrt{2a}(b-\chi b_y)$. The squared Fourier amplitudes of the bonds give $2a(1+\chi\cos k_x)$ and $2a(1-\chi\cos k_y)$ in the loss matrix. At an edge restrict each original jump to its retained sites. A crossing bond becomes a one-site loss, so the open loss matrix is a positive principal submatrix with unchanged diagonal. Removing the entire crossing jump would give a different boundary problem.

\emph{Which observable?}---For a block of dimension $m$ the equal-mixture survival is $m^{-1}\Tr(UU^\dagger)$. Expanding both exponentials and using cyclicity to cancel the commutator terms gives
\begin{equation}
 S(t)=1-2\langle\Gamma\rangle_m t+2\langle\Gamma^2\rangle_m t^2+O(t^3),
\end{equation}
where $\langle A\rangle_m=\Tr A/m$. At matched mean loss the polarization curvature thus detects the difference of the loss variances. This local-in-time identity does not assert that a general bath produces an equal-linewidth spectrum. For vacuum loss the normally ordered covariance satisfies $C(t)=U C(0)U^\dagger$. The traceless block $K=v\sigma_x-i\delta\sigma_z$ obeys $K^2=\Omega^2\Id$. Its exponential is $\cos(\Omega t)\Id-iK\sin(\Omega t)/\Omega$, with the ratio tending to $t$ at threshold. Taking its squared Frobenius norm gives the survival formula and the short-time polarization $4\lambda^2d_0d_1 f_dt^2+O(t^4)$. With paired thermal jumps of strengths $n_B+1$ and $n_B$ times the original rates, the noise adds $2n_B\Gamma$ to the covariance equation. The stationary background is $n_B\Id$ and the excess propagates homogeneously. Unequal channel occupations would require a separate noise calculation.

For the strip, diagonal similarity by $S_{nn}=(T_L/T_R)^{n/2}$ reduces the two hoppings to $\sqrt{T_RT_L}$ for positive $T_R,T_L$. Right eigenvectors inherit $S$ but the eigenvalues do not. The lower-left cofactor of the inverse problem gives $G_{L1}=T_L^{L-1}/\det(\omega_d\Id-D)$, whereas $G_{1L}$ has $T_R$ in its numerator. The shared denominator cancels only from the polarization, not from either absolute intensity. The Supplemental Material gives the extended symmetry derivations, numerical parameters and verification procedure.

\end{document}